\documentclass[12pt]{article}

\usepackage{graphicx}
\usepackage{dcolumn}
\usepackage{bm}

\newcommand{\nn}{\nonumber}
\newcommand{\bi}{\bibitem}
\newcommand{\be}{\begin{eqnarray}}
\newcommand{\ee}{\end{eqnarray}}
\catcode`\@=11
\def\lsim{\mathrel{\mathpalette\@versim<}}
\def\gsim{\mathrel{\mathpalette\@versim>}}
\def\@versim#1#2{\vcenter{\offinterlineskip
\ialign{$\m@th#1\hfil##\hfil$\crcr#2\crcr\sim\crcr } }}
\catcode`\@=12

\begin{document}

\begin{flushright}
KANAZAWA-03-05\\
\end{flushright}

\begin{center}
{\Large\bf The Flavor Symmetry
\footnote{The published version contains
wrong statements about CP violations.
They are removed in the present version, and
the corresponding errors are corrected.}}
\end{center} 

\vspace{1cm}
\begin{center} 
{\sc Jisuke Kubo} $ ^{(a)}$,
{\sc Alfonso Mondrag\' on}$ ^{(b)}$,\\
{\sc Myriam Mondrag\'on}$ ^{(b)}$ and
{\sc Ezequiel  Rodr\' iguez-J\' auregui}$ ^{(b)}$
\vspace{0.2cm}

${}^{(a)}$Institute for Theoretical Physics, 
Kanazawa  University, 
Kanazawa 920-1192, Japan\\

${}^{(b)}$Instituto de F\' isica, UNAM, Apdo. Postal 20-364, M\'exico
  01000 D.F., M\' exico
\end{center}

\vspace{1cm}
\begin{center}
{\sc\large Abstract}
\end{center}

\noindent
Assuming that 
the lepton, quark and Higgs fields belong to the three-dimensional
reducible representation of the
permutation group $S_3$, we suggest
a minimal $S_3$ invariant extension of the
standard model. We find that in the leptonic sector,
the exact $S_3\times Z_2$ symmetry,
which allows $6$ real independent parameters
with two CP-violating phases,  is consistent with 
experimental data and predicts 
a maximal mixing of the atmospheric neutrinos, and that 
the third neutrino is the lightest neutrino.
With the exact $S_3$ only, 
there are $10$ real  parameters and five phases
 in the quark sector.
 A set of values of these
parameters that are consistent with  experimental 
observations is given.

\vspace{1cm}

\newpage
\section{Introduction}
A non-abelian flavor 
symmetry would explain various phenomena
in flavor physics that appear to be independent at present.
Moreover, this would provide useful 
hints about physics beyond the standard
model (SM). In this paper we argue that there exists
such a  symmetry at the Fermi scale.
This symmetry is the permutation symmetry 
$S_3$\cite{pakvasa1,harari},
which is the smallest non-abelian symmetry.
It is the symmetry of an equilateral triangle, and has a simple geometrical interpretation.

The product groups $S_3\times S_3$ and $S_3\times S_3\times S_3$ have been considered
by many authors in the past to explain the hierarchical structure of
the fermionic matter in the SM\cite{koide,bimax}.  The introduction of
the product groups indeed has proven to be successful.\footnote{See,
  for instance, Ref. \cite{fritzsch3}.} However, these symmetries are
explicitly and hardly broken at the Fermi scale.  
If we accept $S_3$ as a
fundamental symmetry in the matter sector of the SM, we are
automatically led to extend the Higgs sector of the SM, because the SM
contains only one Higgs $SU(2)_L$ doublet, which can only be an $S_3$
singlet: Since $S_3$ has two irreducible representations
\footnote{By this we mean that the three-dimensional representation
of $S_3$, for which the permutations are explicit,
can be decomposed
into the direct sum of two irreducible representations.}, singlet and
doublet, there is no convincing reason why there should exist only an
$S_3$ singlet Higgs.  In fact, along this line of thought, interesting
models based on $S_3, S_4$ and also $A_4$ have been considered
\cite{pakvasa1,yamanaka,koide1,ma,babu}. However, the equality of the
irreducible representations has not been stressed
in these works.  The permutation symmetry
$S_3$ means the equality not only of three objects, but also of its
irreducible representations. Nevertheless, it allows differences among
the generations that are realized in the nature of elementary
particles, as we will see.

\section{A minimal $S_3$ invariant extension of the  standard model}
Consider a set of three objects, $(f_1,f_2,f_3)$, and  their six
possible permutations.
They are the elements of 
$S_3$, which is the discrete non-abelian group with the smallest number
of elements. 
The three-dimensional representation 
 is not an irreducible
representation of $S_3$. It can be decomposed
into the direct sum of two irreducible representations,
a doublet ${\bf f}_D$ and a singlet ${\bf f}_S$,
where
\be 
{\bf f}_S &=& \frac{1}{\sqrt{3}}(f_1+f_2+f_3)~,~
{\bf f}^T_D=(~\frac{1}{\sqrt{2}}(f_1-f_2),
\frac{1}{\sqrt{6}}(f_1+f_2-2f_3)~).
\ee
Two-dimensional matrix representations,
$D_i$,  of $S_3$ can be obtained from
\be
D_+(\theta) &=& \left( \begin{array}{cc}
\cos\theta &\sin\theta   \\ -\sin\theta & \cos\theta   \\
\end{array}\right)~~\mbox{and}~~
D_- (\theta) = \left( \begin{array}{cc}
-\cos\theta &\sin\theta   \\ \sin\theta & \cos\theta   \\
\end{array}\right)
\ee
with $\theta=0,\pm 2\pi/3$, where 
$\det D_{\pm} =\pm 1$. The angles $\theta$'s
correspond to the symmetry of an equilateral triangle.
The tensor product of two doublets, ${\bf p}_D^T
=(p_{D1}, p_{D2})$ and $ {\bf q}_D=(q_{D1}, q_{D2})$,
contain two singlets, ${\bf r}_S$ and ${\bf r}_{S'}$, and
one doublet, $ {\bf r}_D=(r_{D1}, r_{D2}) $, where
\be
{\bf r}_S &=& p_{D1} q_{D1}+p_{D2} q_{D2}~,~
{\bf r}_{S'} = p_{D1} q_{D2}-p_{D2} q_{D1}, \\
{\bf r}_D^T &=&
(r_{D1},r_{D2})
=(p_{D1} q_{D2}+p_{D2} q_{D1},
p_{D1} q_{D1}-p_{D2} q_{D2}). 
\label{doublet}
\ee
Note that ${\bf r}_{S'}$ is not an $S_3$ invariant,
while ${\bf r}_{S}$ is.

After the short description of $S_3$ given above,
it is straightforward to extend the SM:
In addition to the SM Higgs fields
$H_S$, we introduce  an $S_3$-doublet Higgs $H_D$.
The quark, lepton and Higgs fields are 
\be
Q^T=(u_L,d_L) &, & u_R~,~d_R~,~L^T=(\nu_L,e_L)~,~e_R~,~ 
\nu_R~,~H
\ee
with obvious notation.
All of these fields have three species, and  we assume that 
each forms
a reducible representation ${\bf 1}_S+{\bf 2}$.
The doublets carry capital indices $I$ and $J$, which run from $1$ to $2$,
and the singlets are denoted by
$Q_3 , u_{3R},d_{3R},L_3~,~
e_{3R}~,~\nu_{3R}~,~H_S$.
Note that the subscript
$3$ has nothing to do with the third generation.
The most general renormalizable Yukawa interactions are given by
\be
{\cal L}_Y &=& {\cal L}_{Y_D}+{\cal L}_{Y_U}
+{\cal L}_{Y_E}+{\cal L}_{Y_\nu},
\ee
where
\be
{\cal L}_{Y_D} &=&
- Y_1^d \overline{ Q}_I H_S d_{IR} - Y_3^d \overline{ Q}_3 H_S d_{3R}  \nn\\
&  &   -Y^{d}_{2}[~ \overline{ Q}_{I} \kappa_{IJ} H_1  d_{JR}
+\overline{ Q}_{I} \eta_{IJ} H_2  d_{JR}~]\nn\\
&  & -Y^d_{4} \overline{ Q}_3 H_I  d_{IR} - Y^d_{5} \overline{ Q}_I H_I d_{3R} 
+~\mbox{h.c.} ,
\label{lagd}\\
{\cal L}_{Y_U} &=&
-Y^u_1 \overline{ Q}_{I}(i \sigma_2) H_S^* u_{IR} 
-Y^u_3\overline{ Q}_3(i \sigma_2) H_S^* u_{3R} \nn\\
&  &   -Y^{u}_{2}[~ \overline{ Q}_{I} \kappa_{IJ} (i \sigma_2)H_1^*  u_{JR}
+\eta  \overline{ Q}_{I} \eta_{IJ}(i \sigma_2) H_2^*  u_{JR}~]\nn\\
&  &
-Y^u_{4} \overline{ Q}_{3} (i \sigma_2)H_I^* u_{IR} 
-Y^u_{5}\overline{ Q}_I (i \sigma_2)H_I^*  u_{3R} +~\mbox{h.c.},
\label{lagu}
\\
{\cal L}_{Y_E} &=& -Y^e_1\overline{ L}_I H_S e_{IR} 
-Y^e_3 \overline{ L}_3 H_S e_{3R} \nn\\
&  &  - Y^{e}_{2}[~ \overline{ L}_{I}\kappa_{IJ}H_1  e_{JR}
+\overline{ L}_{I} \eta_{IJ} H_2  e_{JR}~]\nn\\
 &  & -Y^e_{4}\overline{ L}_3 H_I e_{IR} 
-Y^e_{5} \overline{ L}_I H_I e_{3R} +~\mbox{h.c.},
\label{lage}\\
{\cal L}_{Y_\nu} &=& -Y^{\nu}_1\overline{ L}_I (i \sigma_2)H_S^* \nu_{IR} 
-Y^\nu_3 \overline{ L}_3(i \sigma_2) H_S^* \nu_{3R} \nn\\
&  &   -Y^{\nu}_{2}[~\overline{ L}_{I}\kappa_{IJ}(i \sigma_2)H_1^*  \nu_{JR}
+ \overline{ L}_{I} \eta_{IJ}(i \sigma_2) H_2^*  \nu_{JR}~]\nn\\
 &  & -Y^\nu_{4}\overline{ L}_3(i \sigma_2) H_I^* \nu_{IR} 
-Y^\nu_{5} \overline{ L}_I (i \sigma_2)H_I^* \nu_{3R}+~\mbox{h.c.},
\label{lagnu}
\ee
and
\be
\kappa &=& \left( \begin{array}{cc}
0& 1\\ 1 & 0\\
\end{array}\right)~~\mbox{and}~~
\eta = \left( \begin{array}{cc}
1& 0\\ 0 & -1\\
\end{array}\right).
\label{kappa}
\ee
Furthermore, we introduce the Majorana mass terms for the right-handed
neutrinos 
\be
{\cal L}_{M} &=& 
-M_1 \nu_{IR}^T C \nu_{IR} 
-M_3 \nu_{3R}^T C \nu_{3R},
\label{majo}
\ee
where $C$ is the charge conjugation matrix.

Because of the presence of three Higgs fields,
the Higgs potential $V_H(H_S,H_D)$ is more complicated than that
of the SM. But we may assume 
that all the VEV's are real and that
$\langle H_{1} \rangle = \langle H_{2} \rangle $.\footnote{See, for instance,
Ref. \cite{koide1} in which a potential with three Higgs fields 
of $S_3$ is considered.}
They also satisfy the constraint 
$ \langle H_S \rangle^2
+ \langle H_{1}\rangle^2+ \langle H_{2} \rangle^2
\simeq (246~\mbox{ GeV})^2/2$.
Then from the Yukawa interactions 
(\ref{lagd})--(\ref{lagnu}) and (\ref{majo})
one derives the mass matrices, which have the
general form
\be
{\bf M} = \left( \begin{array}{ccc}
m_1+m_{2} & m_{2} & m_{5} 
\\  m_{2} & m_1-m_{2} &m_{5}
  \\ m_{4} & m_{4}&  m_3
\end{array}\right).
\label{general-m}
\ee
The  Majorana masses for $\nu_L$ can be obtained from 
the see-saw mechanism\cite{yanagida},
and the corresponding mass matrix is given by $
{\bf M_{\nu}} = {\bf M_{\nu_D}}\tilde{{\bf M}}^{-1} 
({\bf M_{\nu_D}})^T$,
where $\tilde{{\bf M}}=\mbox{diag}(M_1,M_1,M_3)$.
All the entries in the mass matrices can be complex;
there is no restriction coming from $S_3$.
Therefore, there are $4 \times 5=20$ 
complex parameters in the mass
matrices, which should be compared with $4 \times 9=36 $
of the SM with the Majorana masses of the left-handed neutrinos.
The mass matrices are diagonalized by the unitary matrices
as
\be
U_{d(u,e)L}^{\dag}{\bf M}_{d(u,e)}U_{d(u,e)R} 
&=&\mbox{diag} (m_{d(u,e)}, m_{s(c,\mu)},m_{b(t,\tau)}),
\label{UeL}\\
U_{\nu}^{T}{\bf M_\nu}U_{\nu} &=&
\mbox{diag} (m_{\nu_1},m_{\nu_2},m_{\nu_3}).
\label{Unu}
\ee
The mixing matrices are then defined as
\footnote{We denote the 
physical neutrino  masses by $m_{\nu_i}$, 
but $\nu_{iL}$ are not the mass eigenstates.}
\be
V_{CKM} &=&U_{uL}^{\dag} U_{dL}~,~
V_{MNS} =U_{eL}^{\dag} U_{\nu}.
\label{ckm1}
\ee

\section{The leptonic sector and $Z_2$ symmetry}
To achieve  further reduction of the number of
parameters, 
we  introduce a  $Z_2$ symmetry.
The $Z_2$ assignment in the leptonic sector is given in the Table I.
\begin{center}
\footnotesize
{\bf Table I}. $Z_2$ assignment in the leptonic sector.

\normalsize
\begin{tabular}{|c|c|}
\hline
 $-$ &  $+$
\\ \hline

$H_S, ~\nu_{3R}$ & $H_I, ~L_3, ~L_I, ~e_{3R},~ e_{IR},~\nu_{IR}$
\\ \hline
\end{tabular}
\end{center}
The $Z_2$ symmetry forbids certain couplings:
\be
 Y^e_{1} &=& Y^e_{3}= Y^{\nu}_{1}= Y^{\nu}_{5}=0.
\label{zeros}
\ee
[The $Z_2$ assignment above  is not the unique assignment   to achieve
(\ref{zeros}).]
Since $m_1^e=m_3^e=0$ due to the $Z_2$ symmetry, 
all the phases appearing
in (\ref{general-m}) can be removed by
a redefinition of $L_I,L_3$ and $e_{3R}$.
Then, we calculate the unitary matrix $U_{eL}$  from
\be
U_{eL}^{\dag} {\bf M}_e{\bf M}_e^{\dag} U_{eL} =
\mbox{diag} (|m_e|^2,|m_\mu|^2,|m_\tau|^2),
\ee
where
\be
{\bf M}_e{\bf M}_e^{\dag}
&=&\left( \begin{array}{ccc}
2 (m^e_{2})^2+(m_{5}^e)^2 &  
(m_{5}^e)^2&  2 m^e_{2} m_{4}^e
\\  (m_{5}^e)^2 &2 (m^e_{2})^2
+(m_{5}^e)^2& 0
  \\  2 m^e_{2} m_{4}^e & 0 &
 2 (m_{4}^e)^2
\end{array}\right).
\label{ml2}
\ee
All the parameters in (\ref{ml2}) are real.
The Majorana masses of the right-handed neutrinos
  $M_{1}$ and $M_{3}$ in (\ref{majo}) may be complex.
They can be made real by a redefinition of the right-handed neutrino fields.
Then  we redefine $m_i^\nu$ according to
\be
(m_2^\nu) &\to & \rho_2^\nu =(m_2^\nu) /\sqrt{M_1}~,~
(m_4^\nu) \to \rho_4^\nu = (m_4^\nu) /\sqrt{M_1}~,~
(m_3^\nu) \to  \rho_3^\nu =(m_3^\nu) /\sqrt{M_3}.
\label{rescale}
\ee
to obtain
the  Majorana masses of the left-handed neutrinos 
in  the form
\be
{\bf M}_{\nu} & = &{\bf M_{\nu_D}}\tilde{{\bf M}}^{-1} 
({\bf M_{\nu_D}})^T=
\left( \begin{array}{ccc}
2 (\rho^{\nu}_{2})^2 & 0 & 
2 \rho^{\nu}_2 \rho^{\nu}_{4}
\\ 0 & 2 (\rho^{\nu}_{2})^2 & 0
  \\ 2 \rho^{\nu}_2 \rho^{\nu}_{4} & 0  &  
2 (\rho^{\nu}_{4})^2 +
(\rho^{\nu}_3)^2
\end{array}\right).
\label{m-nu}
\ee
All the phases in (\ref{m-nu}), 
except for e.g.  $\rho_3^\nu $,  can  be  eliminated.
However, in the following discussions,
we assume that   $(\rho_3^\nu)^2 $ is real,
so that $\rho_3^\nu $ is either a real or a purely imaginary
number.

Now, consider the limit $ m_{4}^e
\to 0$ in (\ref{ml2}). One of the eigenvalues of (\ref{ml2})  becomes $0$. 
Therefore,  we assume
that $m_e^2 \sim  (m_{4}^e)^2$.
In this limit, the other eigenvalues,
$(m_\mu^2~,~m_\tau^2)$, and
the corresponding eigenvectors, 
${\bf v}_\mu$ and ${\bf v}_\tau$, are given by
\be
(m_\mu^2~,~m_\tau^2)&= &
(2( m^e_{2})^2~,~2( m^e_{2})^2
+2( m^e_{5})^2),\\
{\bf v}_\mu &=& (-1/\sqrt{2}~,~1/\sqrt{2})~,~
 {\bf v}_\tau= (1/\sqrt{2}~,~1/\sqrt{2}).
\ee
Therefore,  $U_{eL}$ in this limit becomes
\be
U_{eL}^0 &= &\left( \begin{array}{ccc}
0 & 1/\sqrt{2} &  1/\sqrt{2}
\\  0 & -1/\sqrt{2}  & 1/\sqrt{2}
  \\ 1 & 0 &  0
\end{array}\right).
\label{UeL2}
\ee
The correction to
the eigenvalues due to the nonvanishing $ m^e_{5}$
can be computed, and we find
\be
m_e^2 &=& \frac{(m_{4}^e 
m_{5}^e)^2}{(m^e_{2})^2+(m_{5}^e)^2}
+O((m_{4}^e)^4),\\
~m_\mu^2 &=& 2 (m^e_{2})^2+
(m_{4}^e)^2+O((m_{4}^e)^4),\\
m_{\tau}^2 &=&  2[~(m^e_{2})^2+(m_{5}^e)^2~]+
\frac{(m_{4}^e m^e_2)^2}{(m^e_2)^2
+(m_{5}^e)^2}+O((m_{4}^e)^4).
\ee
For the mass values
$m_e=0.51$ MeV, $m_\mu =105.7$ MeV and $m_\tau
=1777$ MeV (which correspond to
$m_4^e/m_2^e=0.006836$ and $ m_5^e/m_2^e=16.78$),  we obtain
\be
U_{eL} &\simeq &\left( \begin{array}{ccc}
 3.4 \times 10^{-3} &
 1/\sqrt{2} +O(10^{-5})&  1/\sqrt{2} +O(10^{-10})
\\ -3.4 \times 10^{-3}&  -1/\sqrt{2}  +O(10^{-5}) & 1/\sqrt{2} +O(10^{-10}) 
\\    -1+O(10^{-5})  & 4.8 \times 10^{-3} & O( 10^{-5} ) 
 \end{array}\right).
\label{UeL3}
\ee

In the neutrino sector, 
one immediately finds that one of the eigenvalues of
${\bf M}_\nu $ is $2 (\rho^\nu_{2})^2$ with the eigenvector
$(0, 1,0)$. [Recall that $(2 \rho^\nu_{3})^2$ is assumed to be real.]
The other eigenvalues $m_\pm$ are
\be
m_\pm
&=&
\frac{1}{2}(A\pm [-8 (\rho^\nu_{2}
 \rho_{3}^\nu )^2+A^2]^{1/2}),
\label{mpm}
\ee
where
$A =2 (\rho^\nu_{2})^2+(\rho_{3}^\nu)^2
+2 (\rho_{4}^\nu)^2$.
Noticing that
\be
\tilde{A} &=& -8 (\rho^\nu_{2}
 \rho_{3}^\nu )^2+A^2 = (-2(\rho^\nu_{2})^2+
 (\rho^\nu_{3})^2)^2+4 (\rho^\nu_{4})^2
 ((\rho^\nu_{4})^2+2(\rho^\nu_{2})^2+(\rho^\nu_{3})^2),\nn
 \ee
we find that $m_-$ ($m_+$) reaches the maximal 
 (minimal) value,
$2(\rho^\nu_{2})^2$, at $\tilde{A}=0$,
if $(\rho^\nu_{3})^2$ is positive.
For a positive $(\rho^\nu_{3})^2$,
we can obtain $\tilde{A}=0$ 
if $2 (\rho^\nu_{2})^2=(\rho_{3}^\nu)^2$ 
and $(\rho_4^\nu)^2=0$ are satisfied.
Therefore,
\be
m_- & \leq 2 (\rho^\nu_{2})^2 \leq m_+
\ee
should be satisfied if $(\rho_{3}^\nu)^2$ is positive.
Consequently, if $\rho_{3}^\nu$ is  real,
$2 (\rho^\nu_{2})^2 $ cannot be identified with
$ m_{\nu_3}$, which comes from  the experimental constraint
$|\Delta m_{12}^2| <<  |\Delta m_{23}^2|$. 
Thus,  we have to
identify it with $ m_{\nu_1}$. Consequently,  we arrive at
\be
(U_{\nu})_{2 1}  &=& 1, (U_{\nu})_{1 1}=(U_{\nu})_{3 1}=0,
\ee
which does not yield the  experimentally
preferred bi-maximal form of the mixing
matrix $V_{\rm MNS}=U_{eL}^{\dag} U_{\nu}$ with $U_{eL}$  given in 
(\ref{UeL3}). The bi-maximal form\cite{bimax}
may be obtained if  $U_{\nu}$ takes the form
\be
U_{\nu}^{\rm max} &= &\left( \begin{array}{ccc}
1/\sqrt{2} & 1/\sqrt{2}  & 0\\
0 & 0  &  1
  \\  -1/\sqrt{2} &  1/\sqrt{2} &  0
\end{array}\right).
\label{unumax}
\ee
Therefore, to realize the bi-maximal form,
$2 (\rho^\nu_{2})^2 $ has to be the smallest
eigenvalue of the mass matrix (\ref{m-nu}). This, however,
is impossible if $(\rho_{3}^\nu)^2   \geq 0$, implying that
 $\rho_{3}^\nu$ has to be purely imaginary
 (under the assumption that $(\rho_3^\nu)^2$ is real).
  Therefore we write
 $\rho_{3}^\nu$ as
\be
(\rho_{3}^\nu)^2 &=&-|\rho_{3}^\nu|^2.
\ee
As a consequence, the mass relation
$ 2(\rho^\nu_2)^2   <   |m_-|,  |m_+|$
is realized, so  that
the third neutrino becomes the lightest neutrino
with the mass
\be
m_{\nu_3}  &= & 2 (\rho^\nu_2)^2.
\ee
This is  one of the important
predictions of the $S_3\times Z_2$ symmetry,
under the assumption that $(\rho_3^\nu)^2$ is real.

Now consider the limit $ \rho^\nu_{2} \to 0$
with the constraint $(M_\nu)_{33} =0$,
that is $2 (\rho_{4}^\nu)^2-|\rho_{3}^\nu|^2=0$.
Then the eigenvalues  are given by
\be
& &\left(~m_+=2 | \rho_{4}^\nu \rho^\nu_{2}|+(\rho^\nu_{2})^2
~,~m_-= -2 | \rho_{4}^\nu \rho^\nu_{2}|+ (\rho^\nu_{2})^2~,
~m_{\nu_3}=2 (\rho^\nu_{2})^2
~\right),
\label{eigen3}
\ee
and 
\be
U_\nu & \to &U_\nu^{\rm max},\nn
\ee
where $U_\nu^{\rm max}$
is given in (\ref{unumax}), and   $m_{\nu_1}=|m_-|$
and $ m_{\nu_2}=|m_+|$.
So, the limiting form is exactly  the bi-maximal form.

The closed form for $U_\nu$ is found to be
\be
U_{\nu}&= &\left( \begin{array}{ccc}
\sin\hat{\theta} & \cos\hat{\theta}
&  0
\\ 0 & 0 &1
\\    -\cos\hat{\theta}  & \sin\hat{\theta} & 0
 \end{array}\right),
\label{unumax3}
\ee
with
\be
\tan\hat{\theta}
& =&\left(\frac{m_{+(-)}-m_{\nu_3}}{m_{\nu_3}-m_{-(+)}}\right)^{1/2}~
~\mbox{for}~ |m_+| > (<)  |m_-|,
\ee
where $m_\pm$ is given in (\ref{mpm})  with
the replacement $(\rho_3^\nu)^2 \to -|\rho_3^\nu|^2$, and 
$m_{\nu_2}=|m_{+(-)}|$ and $m_{\nu_1}=|m_{-(+)}|$
for $|m_+| > (<)  |m_-|$. Then, together with
(\ref{UeL3}), we obtain
\be
|V_{\rm MNS}| &=&|U_{eL}^{\dag}P U_{\nu}|\nn\\
&\simeq&
\left( \begin{array}{ccc}
\cos\theta_{\rm sol}  & \sin\theta_{\rm sol} & |U_{e3}|
\\  \sin\theta_{\rm sol}/\sqrt{2} 
 & \cos\theta_{\rm sol}/\sqrt{2} 
&   \sin\theta_{\rm atm}
  \\  \sin\theta_{\rm sol}/\sqrt{2}
&  \cos\theta_{\rm sol} /\sqrt{2}
 & \cos\theta_{\rm atm} \end{array}\right),
\label{vmix2}\nn\\
\ee
where $P=\mbox{diag.}(1,1,
\exp i [\arg(Y_4^\nu)-\arg(Y_2^\nu)+\arg(Y_2^e)-
\arg(Y_4^e)]$,
\be
\tan\theta_{\rm atm} &=& 1,
\label{tatm}\\
\tan\theta_{\rm sol} &=&\tan\hat{\theta},\\
 |U_{e3}| &= & 3.4 \times 10^{-3}+O(m_e m_\mu/m_\tau^2).
\label{ue3}
\ee
[Similar, but different predictions have been made in Ref. \cite{fukuyama}.] The results above 
(\ref{vmix2})-(\ref{ue3}) do not change
even if the assumption on the reality of 
$(\rho_3^\nu)^2$ is removed.
In Fig. 1,  we plot $\tan\theta_{\rm sol}$ as a function
of $x=|\Delta m_{23}^2|/m_{\nu_2}^2$ for 
 $r=|\Delta m_{12}^2|/|\Delta m_{23}^2|=0.05~(\mbox{dashed}),\\
0.0264 ~(\mbox{solid} ), 0.2 ~(\mbox{dot-dashed} )$, 
where  $|\Delta m_{23}^2|=m_{\nu_2}^2-
m_{\nu_3}^2$ and $|\Delta m_{12}^2|=
m_{\nu_2}^2-m_{\nu_1}^2$. Note that in the present
model $m_{\nu_3}< m_{\nu_2}, m_{\nu_1}$.
In  Ref. \cite{pakvasa2} (see the  references therein 
and also Ref. \cite{maltoni}), 
 the experimental data  recently
obtained in different neutrino experiments,
including solar\cite{exps-s}, atmospheric\cite{exp-atm},
accelerator neutrino\cite{exp-ac} and reactor\cite{exp-rt} experiments
are reviewed.  It is  concluded that
\be
\tan\theta_{\rm atm} &=& 0.65-1.5~,~
\tan\theta_{\rm sol} = 0.53-0.93,\nn\\
|\Delta m_{12}^2|/\mbox{eV}^2 &=& 
5.1 \times 10^{-5}-9.7\times 10^{-5} ~\mbox{or}~
1.2 \times 10^{-4}-1.9\times 10^{-4},\nn\\
|\Delta m_{23}^2|/\mbox{eV}^2 &=& 1.2\times 10^{-3}-4.8\times 10^{-3},\nn\\
| U_{e3}| & < & 0.2.\nn
\ee
Comparing Fig.~1, (\ref{tatm}) and (\ref{ue3}) 
with the experimental values above, we see that our prediction based on 
the exact $S_3\times Z_2$ symmetry
in the leptonic sector  is consistent
with the most recent
experimental data on neutrino oscillations and neutrino masses and mixings.

\begin{center}
\begin{figure}[htb]
\includegraphics*[width=0.6\textwidth]{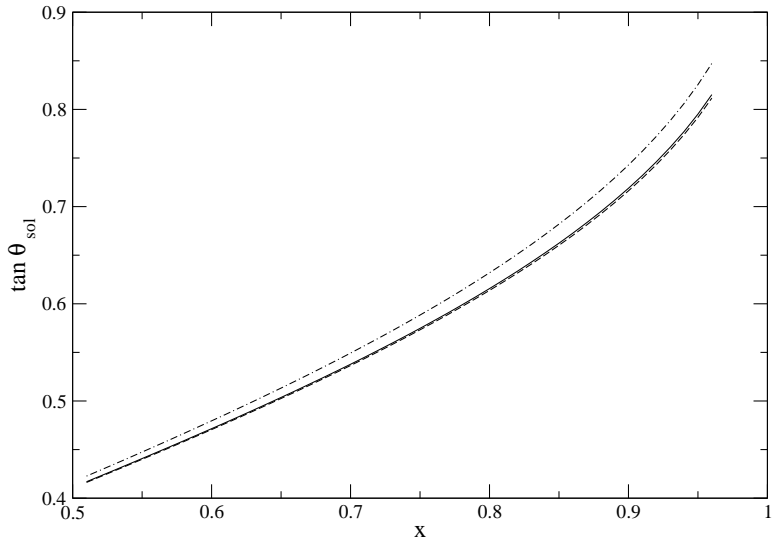}
\caption{$\tan\theta_{\rm sol}$ as a function of
$x=|\Delta m_{23}^2|/m_{\nu_2}^2$ for 
 $r=|\Delta m_{12}^2|/|\Delta m_{23}^2|=0.05 ~(\mbox{dashed}),
0.0264 ~(\mbox{solid} ), 0.2 ~(\mbox{dot-dashed} )$.}
\end{figure}
\end{center}

\section{The hadronic sector}
Now we come to the hadronic sector.
At the level of the $S_3$ extension of the SM, the $Z_2$ assignment
in the hadronic sector  is 
independent of that of the leptonic sector.
(The $Z_2$  assignment in the leptonic sector is given in the Table I.)
Since the $Z_2$ symmetry in the leptonic
sector seems to be  a good symmetry, we assume that it
is a good symmetry at a more fundamental level, too.
Therefore, we require that the $Z_2$ symmetry is free from any quantum 
anomaly.\footnote{Anomalies of discrete symmetries are discussed in
 Ref.  \cite{banks}.}
Furthermore, we assume that the quarks and leptons
are unified at  the fundamental level, which is possible 
if they have the same $Z_2$ assignment, implying that
all the quarks should have  even parity.
One can easily convince oneself that under this 
$Z_2$ assignment the non-abelian gauge anomalies
\be
Z_2[SU(2)_L]^2 &,& Z_2[SU(3)]^2~,~
Z_2[U(1)_Y]^2~,~[Z_2]^3
\ee
cancel at the Fermi scale, because only the right-handed $S_3$
singlet neutrino among fermions has an odd  $Z_2$ parity.

As in the case of  the leptonic sector,
the $Z_2$ symmetry forbids the Yukawa couplings
 $Y^{u,d}_{1}$ and $Y^{u,d}_{3}$ [see (\ref{zeros})].
 We proceed to consider  the 
generation structure in the hadronic sector
under the assumption that $Z_2$ is explicitly  broken in this sector.
Since all the $S_3$ invariant Yukawa couplings are now allowed, the
mass matrices 
for the quarks take the general form (\ref{general-m}),
in which all the entries can be complex.  One can easily see that
all the phases, except for those of 
$m_1^{u,d}$ and $m_3^{u,d}$,
 can be removed  through an appropriate redefinition
of the quark fields. Here, we assume  that
only $m_3^{d}$ is a complex number.
The unitary matrices $U_{uL}$ and  $U_{dL}$ can be obtained from
\be
U_{u(d)L}^{\dag} {\bf M}_{u(d)}{\bf M}_{u(d)}^{\dag} U_{u(d)L} =
\mbox{diag} (|m_{u(d)}|^2,|m_{c(s)}|^2,|m_{t(b)}|^2),
\ee
where
\be
{\bf M}_{u(d)}
&=& \left( \begin{array}{ccc}
m_1^{u(d)}+m_{2}^{u(d)} & m_{2}^{u(d)} & m_{5}^{u(d)} 
\\  m_{2}^{u(d)} & m_1^{u(d)}-m_{2}^{u(d)} &m_{5}^{u(d)}
  \\ m_{4}^{u(d)} & m_{4}^{u(d)}&  m_3^{u(d)}
\end{array}\right).
\label{mud}
\ee
To diagonalize the mass matrices, we start by observing that
realistic mass hierarchies can be achieved in the following way.
In the limit $m_{4,5} \to 0$, they become block-diagonal, and 
$m_3$ becomes an eigenvalue
whose eigenvector is $(0,0,1)$. The $2\times 2$ blocks,
which are of a semi-democratic type, can be simply diagonalized.
One finds easily that  
one of the  eigenvalues can become $0$ if   $m_1^2-2 m_2^2=0$ is satisfied.
So, the gross structure of realistic mass matrices can be obtained if
$m_3^{u,d} \sim O(m_{t,b})$ and  $m_{1,2}^{u,d} \sim O(m_{c,s})$ 
(to realize realistic mass hierarchies), and the non-diagonal elements
$m_{4}^{u,d}$ and $m_{5}^{u,d} $, along   with $m_{1,2}^{u,d}$,
can produce a realistic mixing among the quarks.
There are  $10$ real parameters and one phase
(under the assumption
that only $ m_{3}^{d}$ is complex) in order
to produce  six quark masses, three mixing angles and one 
$CP$-violating phase.
It is certainly desirable to investigate the complete parameter space
of the model
to understand its phenomenology and 
to make predictions, if any can be obtained. However, this is a quite complex problem, and 
will go beyond the scope of
the present paper. Here we would like to give one set of parameters
that are consistent with the experimental values
given by the Particle Data Group\cite{pdg}.
We find that the set
of dimensionless parameter values
\be
m_1^u/m_0^u &=& -0.000293~,~ m_2^u/m_0^u =-0.00028 ~,~   m_3^u/m_0^u=1~,\nn\\  
m_4^u/m_0^u &=&0.031  ~,~m_5^u/m_0^u=0.0386,\nn \\
m_1^d/m_0^d &=&0.0004~,~  m_2^d/m_0^d =0.00275~,~   m_3^d/m_0^d=1+ 1.2I~,\\   
m_4^d/m_0^d  &=& 0.283~,~  m_5^d /m_0^d =0.058\nn
\label{choice1}
\ee
yields the mass hierarchies
\be
m_u/m_t &=&1.33\times 10^{-5}~,~m_c/m_t=2.99\times 10^{-3},\nn\\
m_d/m_b &=&1.31\times 10^{-3}~,~m_s/m_b=1.17\times 10^{-2},
\ee
where $m_0^u=m_3^u$ and $m_0^d=\mbox{Re}(m_3^d)$, and
the mixing matrix becomes
\be
V_{\rm CKM} &=& U_{uL}^{\dag} U_{dL}\nn\\
&=&
\left( \begin{array}{ccc}
0.968+0.117 I & 0.198+0.0974 I &-0.00253-0.00354 I
\\  -0.198+0.0969 I & 0.968-0.115 I  & -0.0222-0.0376 I  \\
 0.00211+0.00648 I &0.0179-0.0395 I & 0.999-0.00206 I
\end{array}\right).\nn\\
\label{vckm1}
\ee
The magnitudes of the elements are given by
\be
|V_{\rm CKM}| &=&
\left( \begin{array}{ccc}
0.975& 0.221 &0.00435
\\  0.221 & 0.974  & 0.0437
  \\ 0.00682 &0.0434 & 0.999
\end{array}\right),
\label{vckm2}
\ee
which should be compared with the experimental values\cite{pdg}
\be
|V_{\rm CKM}^{\rm exp}| &=&
\left( \begin{array}{ccc}
0.9741~\mbox{to}~ 0.9756 & 0.219 ~\mbox{to}~ 0.226 & 0.0025 ~\mbox{to}~ 0.0048
\\  0.219 ~\mbox{to}~ 0.226  & 0.9732 ~\mbox{to}~ 0.9748   & 0.038 ~\mbox{to}~ 0.044
  \\ 0.004 ~\mbox{to}~ 0.014  &0.037 ~\mbox{to}~ 0.044  & 0.9990 ~\mbox{to}~ 0.9993
\end{array}\right).
\ee
Note that the mixing matrix (\ref{vckm1}) is not in the standard parametrization.
So, we give the invariant measure of $CP$-violations\cite{jarskog}
\be
J &=&\mbox{Im}~[(V_{\rm CKM})_{11}(V_{\rm CKM})_{22}
(V_{\rm CKM}^*)_{12}(V_{\rm CKM}^*)_{21}]=2.5\times 10^{-5}
\ee
for the choice (\ref{choice1}), which is slightly larger than  the experimental value
 $(3.0\pm 0.3)\times 10^{-5}$ 
(see \cite{pdg} and also Ref. \cite{b-factory}).
The angles of the unitarity triangle 
for $V_{\rm CKM}$ (\ref{vckm1}) are given by
\be
\phi_1 \simeq 22^\circ~,~
\phi_3 \simeq 38^\circ,
\ee
where the experimental values are
$\phi_1 = 24^\circ\pm 4^\circ$ and 
$\phi_3 = 59^\circ \pm 13^\circ$ \cite{pdg}.
The normalization masses $m_0^u$ and $m_0^d$ are fixed at
\be
m_0^u =174  ~~\mbox{GeV} ~,~m_0^d=1.8~~\mbox{GeV} 
\ee
for $m_t=174$ GeV and $m_b=3$ GeV,
yielding   $m_u \simeq 2.3  $ MeV, $m_c\simeq 0.52$  GeV, 
$m_d \simeq 3.9  $ MeV and $ m_s= 0.035 $ GeV.
Although these values cannot be directly compared with
the running masses, because
our calculation is at the tree level, it is nevertheless
worthwhile to observe how close they are to \cite{fritzsch3}
\be
m_u(M_Z) &=& 0.9 -2.9~~\mbox{MeV}~,
~m_d(M_Z) = 1.8 -5.3~~\mbox{MeV},\nn\\
~m_c(M_Z)  &=& 0.53 -0.68~~\mbox{GeV}~,
~m_s(M_Z) = 0.035 -0.100~~\mbox{GeV},\nn\\
~m_t(M_Z) &=& 168 -180~~\mbox{GeV}~,
~m_b(M_Z) = 2.8 -3.0~~\mbox{GeV}.
\ee

\section{Flavor changing neutral currents (FCNCs)}
In models with more than one  Higgs $SU(2)_L$ doublet,
as in the case of  the present model,
tree-level FCNCs exist in the Higgs sector.
We therefore calculate the flavor changing Yukawa couplings
to the neutral Higgs  fields, $H_S^0$ and 
$H_I^0~(I=1,2)$, where
$H_S^0$ and $ H_I^0$ stand for the neutral Higgs fields of 
the $S_3$-singlet $H_S$ and the $S_3$- doublet $H_I$, respectively.
The actual values of these couplings
depend on the VEV's of the Higgs fields, and hence
 on the Higgs potential, which we do
not consider in the present paper.
 Since  we only would like to estimate
the size of the tree-level FCNCs here, we simply assume
that
\be
\langle H_S^0 \rangle  &=&\langle  H_1^0\rangle = \langle H_2^0\rangle 
\simeq 246/\sqrt{6}~\mbox{GeV}~\simeq 142/\sqrt{2}~\mbox{GeV}.
\ee
Then,  the flavor changing Yukawa couplings
can explicitly be  calculated, because
we know the explicit values of the unitary matrices $U$'s defined in (\ref{UeL}):
\be
{\cal L}_{\rm FCNC} &=&
\left(\overline{E}_{aL} Y_{a b}^{ES} E_{bR}+
\overline{U}_{aL} Y_{a b}^{US} U_{bR}
+\overline{D}_{aL} Y_{a b}^{DS} D_{bR}\right)~H_S^0 +~\mbox{h.c.}\nn\\
&+&
\left(\overline{E}_{aL} Y_{a b}^{E1} E_{bR}+
\overline{U}_{aL} Y_{a b}^{U1} U_{bR}
+\overline{D}_{aL} Y_{a b}^{D1} D_{bR}\right)~H_1^0+~\mbox{h.c.} \nn\\
&+&
\left(\overline{E}_{aL} Y_{a b}^{E2} E_{bR}+
\overline{U}_{aL} Y_{a b}^{U2} U_{bR}
+\overline{D}_{aL} Y_{a b}^{D2} D_{bR}\right)~H_2^0+~\mbox{h.c.}.
\ee
Here the matrices $E$'s, $U$'s and $D$'s stand for the mass eigenstates, 
and
\be
Y^{E1}
&\simeq&\left( \begin{array}{ccc}
-10^{-5} &  
2.6 \times 10^{-6}& -4.2 \times 10^{-5}
\\ -1.1\times 10^{-3} &
5.3 \times 10^{-4}& -8.8\times 10^{-3}
  \\   -1.2\times 10^{-8}
 & 5.3 \times 10^{-4} & -8.8\times 10^{-3}
\end{array}\right),\\
Y^{E2}
&\simeq&\left( \begin{array}{ccc}
5.1\times 10^{-6} &  
-2.6 \times 10^{-6}& 4.2 \times 10^{-5}
\\  1.1\times 10^{-3}  &
5.3 \times 10^{-4}& 8.8\times 10^{-3}
  \\   1.2\times 10^{-8}
 & -5.3 \times 10^{-4} & -8.8\times 10^{-3}
\end{array}\right),\\
Y^{US}
&\simeq&\left( \begin{array}{ccc}
-4.7\times 10^{-4} &  
3.8 \times 10^{-4}& -7.1 \times 10^{-3}
\\ -3.8\times 10^{-4} &
-3.6 \times 10^{-3}& 7.5\times 10^{-2}
  \\   8.8\times 10^{-3}
 & 9.4\times 10^{-2} & -1.7
\end{array}\right),\\
Y^{U1}
&\simeq&\left( \begin{array}{ccc}
8.3\times 10^{-4} &  
1.6\times 10^{-3}& -3.4 \times 10^{-2}
\\ -1.6\times 10^{-3} &
4.9 \times 10^{-3}& -4.1\times 10^{-2}
  \\   4.3\times 10^{-2}
 & -5.1\times 10^{-2} & -4.2\times 10^{-3}
\end{array}\right),\\
Y^{U2}
&\simeq&\left( \begin{array}{ccc}
-3.3\times 10^{-4} &  
-1.9\times 10^{-3}& 4.1 \times 10^{-2}
\\ 1.9\times 10^{-3} &
3.8 \times 10^{-3}& -3.4\times 10^{-2}
  \\   -5.1\times 10^{-2}
 & -4.2\times 10^{-2} & -4.2\times 10^{-3}
\end{array}\right),\\
Y^{DS}
&\simeq&\left( \begin{array}{ccc}
(1.4+0.44 I)\times 10^{-5} &  
(5.6+0.38 I)\times 10^{-5}& -(1.6+1.6 I) \times 10^{-4}
\\ (5.5+0.38 I)\times 10^{-5} &
(2.8-2.1I) \times 10^{-4}& -(1.4+0.16 I)\times 10^{-3}
  \\   -(7.8+8.0 I)\times 10^{-4}
 & -(7.0+0.82 I)\times 10^{-3} &(1.8+2.1 I)\times 10^{-2}
\end{array}\right),\nn\\
\\
Y^{D1}
&\simeq&\left( \begin{array}{ccc}
-(1.1-0.17 I)\times 10^{-4} &  
-(1.6-1.3 I)\times 10^{-4} & (8.0-0.033 I) \times 10^{-4}
\\ -(1.6-1.3 I)\times 10^{-4} &
-(2.4-1.7 I) \times 10^{-4}& (6.1+1.6 I)\times 10^{-4}
  \\   4.0\times 10^{-3}
 & (3.1+0.82 I)\times 10^{-3} &(6.1+7.3 I)\times 10^{-4}
\end{array}\right),\nn\\
\\
Y^{D2}
&\simeq&\left( \begin{array}{ccc}
(6.2-0.71 I)\times 10^{-5} &  
(1.1-1.4 I)\times 10^{-4}& -(6.4-1.6 I) \times 10^{-4}
\\ (1.0-1.3 I)\times 10^{-4} &
-(3.0-2.8I) \times 10^{-4}& (7.5-0.054 I)\times 10^{-4}
  \\   -(3.2+0.80 I)\times 10^{-3}
 & (3.9-0.001 I)\times 10^{-3} &(6.1+7.3 I)\times 10^{-4}
\end{array}\right).\nn\\
\label{dd}
\ee
All the non-diagonal elements are responsible for
tree-level FCNC processes.
The amplitude of the flavor violating process $\mu^- \to e^+ e^- e^-$, for instance,
is proportional to
$(Y^{E1})_{11}(Y^{E1})_{21} \simeq  10^{-8}$.
Then, 
we find that
its branching ratio is estimated to be
\be
B(\mu \to 3 e) &
\sim 10^{-15} (M_W/M_H)^4 < 10^{-12},
\ee
where $M_W$ and $M_H$ are the $W$ boson mass
and Higgs mass, respectively, and the value $10^{-12}$ is
the experimental upper bound. Similarly, we obtain
\be
B(\tau \to 3 \mu) &
\sim 10^{-10} (M_W/M_H)^4 < 10^{-6},\\
B(K^0_L \to  2e) &
\sim 10^{-16} (M_W/M_H)^4 < 10^{-12},\\
B(B^0_S  \to 2\mu) &
\sim 10^{-7} (M_W/M_H)^4 < 10^{-6}.
\ee
Note that because of the three Higgs fields, the imaginary parts of
the $Y$'s contribute to  
$CP$-violating amplitudes, which are not 
taken into account by the phase of the mixing matrix $V_{\rm CKM}$.
Therefore, the  independent phases 
 in the mass matrices (\ref{general-m})
can be,  in principle, measured.
A complete analysis of this problem will go beyond the scope
of the present paper, and we would like to leave this problem
to a future work.

\section{Conclusion}
$S_3$ is a non-abelian permutation group with the smallest number
of elements.  The symmetry $S_{3L}\times S_{3R}$ has been considered by
many authors\cite{koide,bimax,fritzsch3} in the past to explain the
hierarchal structure of the generations in the SM. $S_{3L}\times S_{3R}$ is,
however, explicitly and hardly broken at the Fermi scale.  In the present paper we
considered its diagonal subgroup, while extending the concept of flavor
and generation to the Higgs sector.  Once this is done, there is no
reason that there should exist only an $S_3$ singlet Higgs, and so we
introduced three $SU(2)_L$ Higgs  doublet fields.  The minimal $S_3$
extension of the SM allows a definite structure of the Yukawa
couplings, and we studied its consequences, in particular the mass of the
quarks and leptons, and their mixings.  Although similar ideas have
been proposed previously \cite{pakvasa1,yamanaka,koide1,ma,babu}, none
of the existing treatments is identical to ours: The main differences are the
inclusion of $(S_3-\mbox{doublet})^3$ couplings and the arrangement of
the $S_3$ representations.  We found that in the leptonic sector, an
additional discrete symmetry $Z_2$ can be consistently introduced.
 The theoretical predictions obtained
from the $S_3 \times Z_2$ symmetry are
consistent with the experimental observations made to this time. 
As for the quark sector,  we simply assumed
that $Z_2$ is explicitly broken and analyzed
the quark mass matrices that result only from $S_3$. We found that
they are consistent with experiments.  From these studies we
 hypothesize  that the flavor symmetry, which is exact at the Fermi
scale, is the permutation symmetry $S_3$.  
This flavor symmetry, together  with
the electroweak gauge symmetry, is only spontaneously broken.
The analysis of the mass
matrices in the hadronic sector that we performed in the present
paper, is by no means complete, because we gave only one set of consistent
parameter values.  Also, the analysis of the FCNCs in the
Higgs sector is not complete.  A more complete study will be published
elsewhere.

Supersymmetrization of the model is straightforward.
It will simplify the Higgs sector drastically, and, moreover,
the squared soft scalar masses  will enjoy a certain degree of  degeneracy,
thanks to $S_3$, thereby softening
the supersymmetric flavor problem.

\section*{Acknowledgements}
One of us (JK)  would like to acknowledge
the kind hospitality of the theory group at the 
Institute for Physics, UNAM,  Mexico.
This work is supported by a Grants-in-Aid
for Scientific Research  from
 the Japan Society for the Promotion of Science (JSPS)
 (No. 13135210),
 and by the UNAM grant PAPIIT-IN116202.
 This work was partially conducted by
 way of a grant awarded by the Government of Mexico in the
 Secretariat of Foreign Affairs.

\end{document}